# Linear magneto-resistance versus weak antilocalization effects in Bi$_2$Te$_3$ films

ZhenHua Wang[1] (✉), Liang Yang[1], XiaoTian Zhao[1], ZhiDong Zhang[1] and Xuan P. A. Gao[2] (✉)

[1] Shenyang National Laboratory for Materials Science, Institute of Metal Research, Chinese Academy of Sciences, 72 Wenhua Road, Shenyang 110016, People's Republic of China
[2] Department of Physics, Case Western Reserve University, Cleveland, Ohio 44106, United States

## ABSTRACT

In chalcogenide topological insulator materials, two types of magneto-resistance (MR) effects are widely discussed: a positive MR dip around zero magnetic field associated with the weak antilocalization (WAL) effect and a linear MR effect which generally persists to high fields and high temperatures. We have studied the MR of topological insulator Bi$_2$Te$_3$ films from the metallic to semiconducting transport regime. While in metallic samples, the WAL is difficult to identify due to the smallness of the WAL compared to the samples' conductivity, the sharp WAL dip in the MR is clearly present in the samples with higher resistivity. To correctly account for the low field MR by the quantitative theory of WAL according to the Hikami-Larkin-Nagaoka (HLN) model, we find that the classical (linear) MR effect should be taken into account together with the WAL quantum correction. Otherwise the WAL fitting alone yields an unrealistically large coefficient in the HLN analysis. This work clarifies the WAL and LMR as two distinct effects and offers an explanation for the overly large in the WAL analysis of topological insulators in some literature.

Address correspondence to Zhenhua Wang, zhwang@imr.ac.cn; Xuan P.A. Gao, xuan.gao@case.edu

# 1 Introduction

Topological insulators (TIs) are quantum matters with metallic surface states surrounding an insulating bulk, as a result of band inversion in compounds with large spin-orbit interaction. [1,2] Many chalcogenides such as $Bi_2Se_3$ and $Bi_2Te_3$ have been experimentally confirmed to have these topological surface states with Dirac dispersion through angle-resolved photoelectron spectroscopy (ARPES) [3,4]. In three-dimensional (3D) TIs with single Dirac cone, the π-Berry phase and helical spin-momentum locking suppress the back-scattering and lead to the well-known WAL effect [5,6,7] which is manifested as a dip-like positive MR in low magnetic fields in the absence of any magnetic scattering. The WAL induced positive MR in small perpendicular magnetic fields was found in $Bi_2Se_3$ films and exfoliated microflakes [8-14], $Bi_2Te_3$ film [15], nanoribbons[16, 17] and microflakes [18]. In addition to the WAL effect, another commonly observed MR is the non-saturating linear MR effect [19]. Unlike the low field WAL, this novel linear MR is less understood, although its existence seems ubiquitous [20-26]. Being the two prevailing types of MR observed in the TI, the relation and interplay between WAL and linear MR have attracted increasing attention in research [26, 27]. In most theoretical models that can give rise to a linear MR, the linear MR is considered to be due to a completely different mechanism than the WAL type of quantum correction effect [28-30]. However, it was also suggested if the linear MR could be merely the high field extension of the WAL [26]. In this work, we compare the low field vs. high field magneto-transport in $Bi_2Te_3$ films with different level of metallicity (or resistivity). These films exhibit a classical linear MR effect which is controlled by the resistivity level and the degree of granularity of the sample [31]. We found that the WAL effect is more clearly present in samples with larger resistivity in which the linear MR effect is weak. For samples with increasingly lower resistivity and stronger linear MR, the WAL fitting of low field magneto-conductivity according to the Hikami-Larkin-Nagaoka (HLN) model produces unreasonably large coefficient (or magnitude) of WAL correction. This over-estimate of the WAL correction can be remedied by subtracting the magneto-conductivity associated with linear MR as a background. This finding highlights that care needs to be taken in the analysis of MR data in TIs in which the WAL could be mixed with other MR effects that have classical or semi-classical origin.

# 2 Experimental Method

$Bi_2Te_3$ films were grown by a vapor transport deposition method on semi-insulating Si substrates with size of ~1.5cm×1.5cm in 10% $H_2$/Ar carrier gas. The growth method is described in details in Ref. [31]. Briefly, 99.99% $Bi_2Te_3$ powder was used as precursor and thermally evaporated inside a one inch diameter quartz tube placed in a single zone tube furnace (Lindberg Blue M) set at 500-520°C. Argon (with 10%$H_2$) carrier gas was flown at 40sccm flow rate to transport the evaporated Bi and Te vapor downstream for the deposition of $Bi_2Te_3$ film on the Si substrate which was 14-15cm away from the $Bi_2Te_3$ source (center of furnace). The $Bi_2Te_3$ films studied in this work were typically grown at pressure of 30-50Pa over 5min. They have Bi to Te ratio about 2:3 and thickness around 100nm as confirmed by energy dispersive X-ray spectroscopy and scanning electron microscope (SEM) imaging [31]. Here we focus on discussing the representative magneto-transport data collected on five samples with increasingly semiconducting behavior, noted as S1, S2, S3, S4, and S5 respectively. The electrical behavior (metallic vs. semiconducting) of the $Bi_2Te_3$ films presented here was varied by changing the growth conditions (temperature, pressure). Generally we found that in the pressure and temperature ranges used (500-520°C, 30-50Pa), higher temperature and pressure yielded more metallic samples with higher mobility (e.g. among S1-S5, S1 was grown at 520°C and 50Pa and showed strongest metallic $R(T)$). As grown samples were cut into ~3mm×10mm rectangle shape and silver paste was put onto the $Bi_2Te_3$ film to obtain Ohmic contact. Their transport properties were investigated by Quantum Design Physical Property Measurement System using Hall bar configuration.

# 3 Results and discussion

In order to compare the basic transport behavior of these samples, the temperature dependent resistance $R$ at $B=0$ for S1, S2, S3, S4, and S5 are shown in Fig. 1(a). The resistance is normalized over its value at 300K, to emphasize the temperature dependence. The respective sheet resistance at 300K is 462, 343, 166, 76, and 4800 Ω/square for S1, S2, S3, S4, and S5. The SEM images of S1, S2 and S5 are shown in Fig. 1 (b), (c) and (d). Similar to Ref. [31], sample S1 is composed of the most densely connected micro-plates of $Bi_2Te_3$ and exhibits the strongest metallic behavior in its temperature dependent $R$ among these five samples. This is correlated with its highest low $T$ mobility (~300cm$^2$/Vs at 2K). On the other hand, samples made of more loosely connected micro-plates show more insulating behavior, reflecting their lower mobility (5-100 cm$^2$/Vs at 2K) . The carrier density is around $10^{19}$-$10^{20}$/cm$^3$ [31]. Given the high carrier density in these films, bulk carriers are expected to dominate the transport.

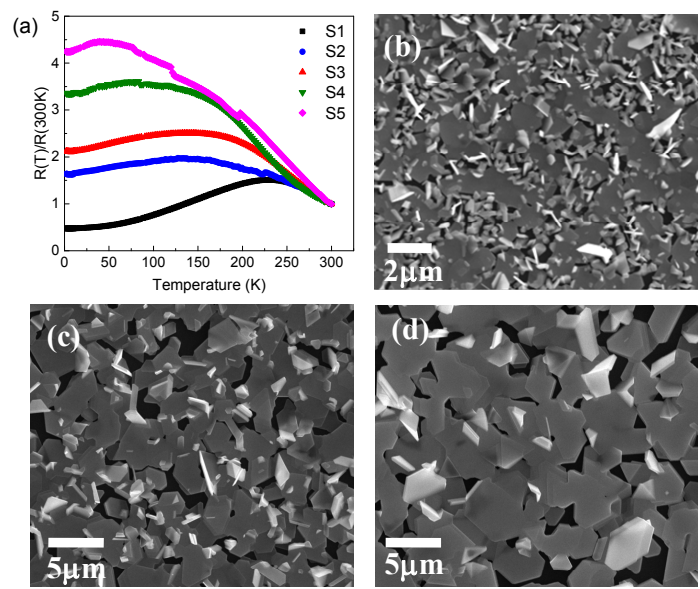

**Figure 1** (a) The temperature dependent resistance, $R(T)/R(T=300K)$ of $Bi_2Te_3$ film S1, S2, S3, S4 and S5. The respective sheet resistance at 300K is 462, 343, 166, 76, and 4800 Ω/square for S1, S2, S3, S4, and S5. (b), (c) and (d) show the SEM images of S1, S2 and S5 respectively.

In our previous work [31], we also observed that these $Bi_2Te_3$ films exhibit mobility fluctuation induced classical linear magneto-resistance whose magnitude correlates with the average mobility of the film and the degree of granularity. Because of higher mobility, more metallic samples generally exhibit stronger MR effect [31]. Such effect is also observed in this study. As can be seen in Fig.2 (a), the perpendicular magnetic field induced MR in metallic sample S1 shows a quadratic behavior at low $B$ and linear MR at high $B$. The MR, ΔR(B)/R(0) reaches nearly 120% from $B=0$ to -5T or +5T and the MR has very little temperature dependence over $T$= 2 - 20K. However, more insulating samples S2, S3, S4, and S5 exhibit much weaker MR effect (<25%) over the same field range, as shown in Fig.2 (b). Comparing Fig.2 (a) and (b), one observes that despite all the films show a linear MR in the high field regime ($B$>1T) [31], the low field ($B$<1T) behavior is different. In S1, the low field MR is quasi-quadratic, similar to the classical MR in metals or semiconductors. However, the MR around $B$=0 appears as a dip in sample S2-S5 (Fig.2(b)) at low temperature. This dip in the MR around $B$=0 is sharpest in the most insulating sample, S5 and becomes progressively weaker from S4, S3 to S2 as the sample gets more metallic in its $R(T)$. This sharp dip in the MR around *B*=0 at low temperatures is the well known WAL effect due to the destructive quantum interference or suppressed back-scattering of carriers [8-18]. The attribution of the MR dip around *B*=0 in S2-S5 is confirmed by its temperature dependence. We plot the MR for S2 and S5 at T=2, 5, 10 and 20K in Fig.2 (c) and (d). As expected in the WAL phenomenon, the MR dip around *B*=0 is sharper at lower T, where the electron dephasing (electron-electron and electron-phonon scattering) is weaker and coherence time is longer. On the other hand, the linear MR effect at *B*>1T does not vary much with temperature, except a shift caused by the *T*-dependence of WAL. This is consistent with the classical nature of the fluctuation or inhomogeniety induced linear MR in disordered materials [29, 31, 32] since the quantum coherence of carriers does not play a role in the origin of linear MR. In the following, we will discuss this WAL induced MR effect in more details and our analysis points to the importance of separating it from other classical MR effects such as the inhomogeneity induced linear MR [29, 31, 32].

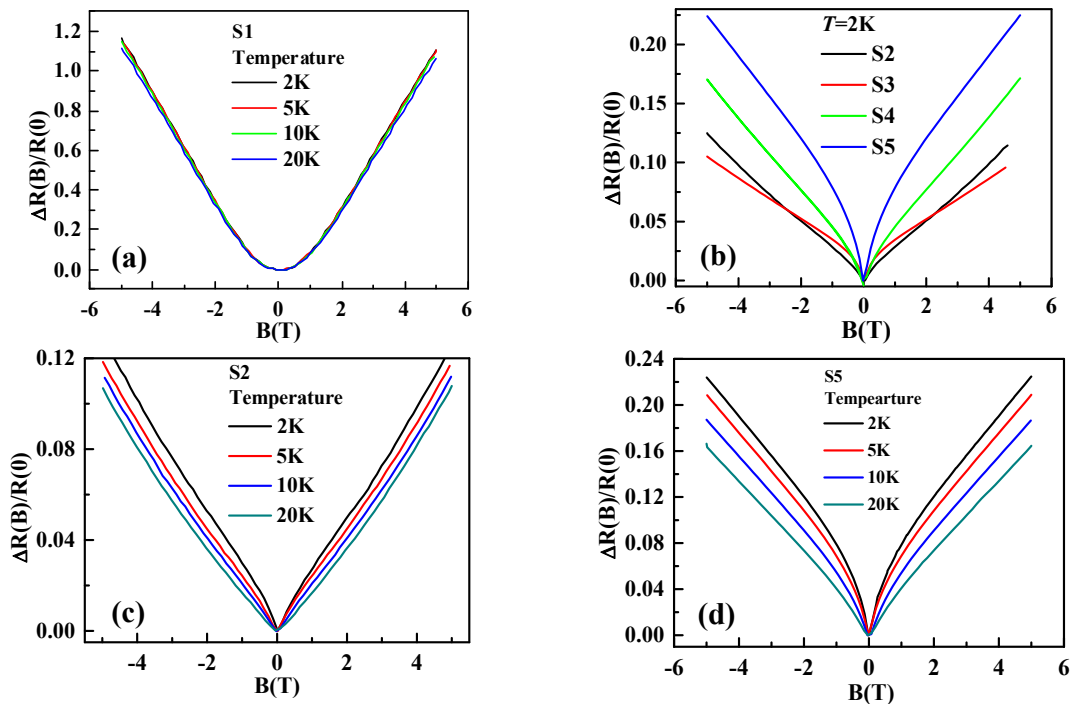

**Figure 2** (a), (c), (d) Magneto-resistance (MR) as a function of the perpendicular magnetic field for Bi2Te3 film sample S1, S2 and S5 at T = 2-20K. (b) MR for S2, S3, S4, and S5 over B= -5T to +5T at T= 2 K.

Since the quantum interference effects give rise to correction in the diffusion constant and thus the conductivity, generally the magneto-conductivity $G_{xx}(B)$ instead of MR is analyzed. We first analyze $G_{xx}(B)$ of sample S5 which exhibits the clearest WAL dip in MR (Fig. 2(d)). Fig. 3(a) shows the $G_{xx}(B)$ in units of $e^2/h$ per square versus perpendicular magnetic field at various temperatures for S5. Here, $G_{xx}$ is obtained by inverting the MR tensor, but because the Hall resistance $R_{xy}$ of these films is much smaller than the longitudinal resistance $R_{xx}$ [31], we have $G_{xx}(B) \approx 1/R_{xx} \times (L/W)$ where *L* is the length and *W* is the width of the sample. After inverting the MR, the WAL effect is shown as a sharp cusp around B=0. At *B*>~1T where the WAL effect is completely destroyed by magnetic field, the magneto-conductivity shows a linearly decreasing trend, as a result of the linear MR. This classical magneto-conductivity effect due to the linear MR is highlighted by a dashed line to guide the eyes for the *T*=2K curve. Fig. 3(b) plots similar magneto-conductivity for sample S2.

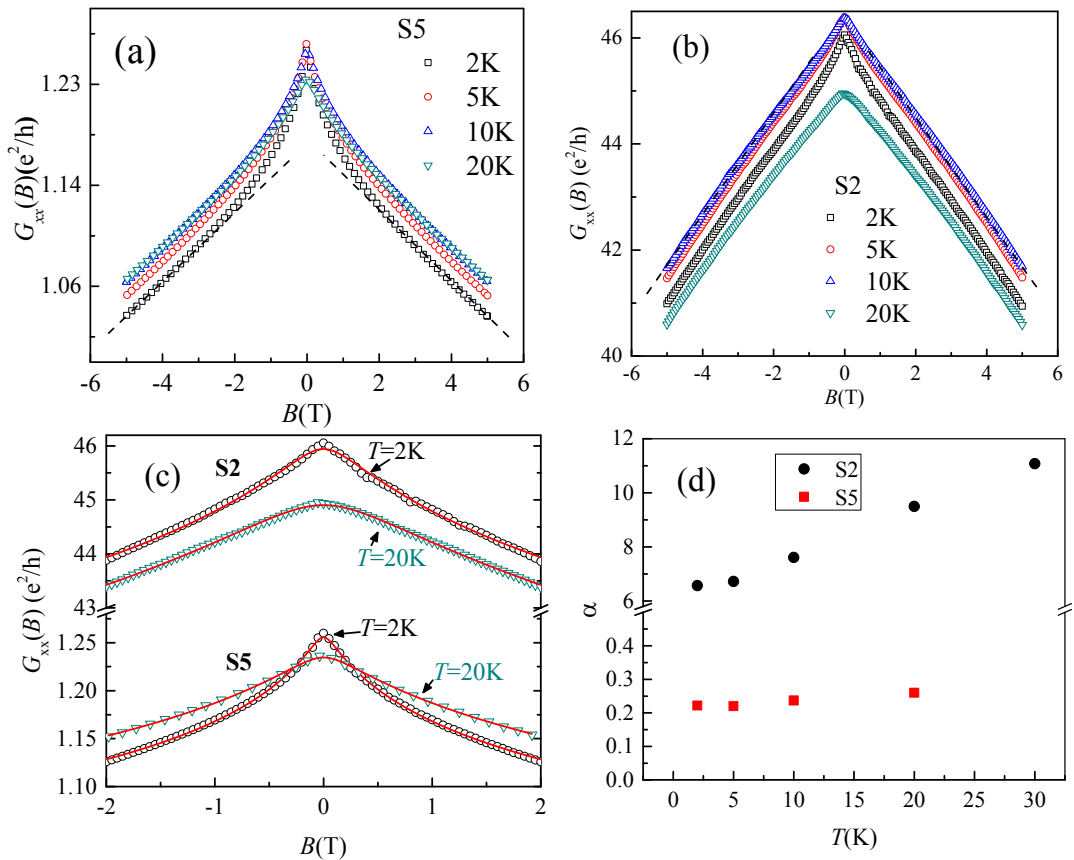

**Figure 3** (a) and (b) show the conductance per square, or 2D conductivity, $G_{xx}$ in units of $e^2/h$ versus perpendicular magnetic field ($B$) at various temperatures for $Bi_2Te_3$ film S5 and S2. (c) shows the direct weak antilocalization (WAL) fitting (solid line) of raw data (dots) for S5 and S2 at $T$=2, 20K; and (d) plots the obtained fitting parameter α as a function of temperature.

The 2D WAL correction to the magneto-conductivity is described by the HLN formula [33]

$$\Delta G_{xx}(B)_{WAL} = \alpha e^2/(2\pi^2\hbar)[\ln(B_\varphi/B) - \psi(1/2 + (B_\varphi/B))] \quad (1),$$

where $B_\varphi = \hbar/(4eL_\varphi^2)$ is a characteristic field defined by the dephasing length $L_\varphi$, ψ is the digamma function and α is a constant [8, 9]. It is known that α=1/2 is expected for a single layer of TI surface states or conventional 2D system with strong spin-orbit scattering [8, 9]. In TI $Bi_2Se_3$ and $Bi_2Te_3$ systems, typical experimental values of α cover the range from 0.1-1, depending on the Fermi level position and thickness of the sample [8-18]. A valued of α=1 was commonly interpreted as the TI film having two independent layers of surface states each contributing α=1/2, while α<1 is generally interpreted as an indication of interlayer or surface-bulk coupling [9, 27]. We first directly fitted the magneto-conductivity data to 2D WAL model (eq.1). A reasonably good fit between the experimental data and eq.1 was obtained as shown in Fig. 3 (c). However, the fitted prefactor α is unrealistically large (α>6) for sample S2 as shown in Fig.3 (d). Similarly, much greater than one were obtained in S3, and S4 (Table 1). We note that compared to S5, samples S2, S3, and S4 all had much higher conductivity value and exhibited large change in the absolute value of conductance as B varies, in association with the classical linear magneto-resistance effect. Without distinguishing WAL from the linear MR effect which has a different origin [31], the large change in $G_{xx}(B)$ due to the classical MR artificially led to an overly large coefficient $\alpha$ in the direct fitting of data to eq. (1).

**Table 1** Comparison of WAL fitting parameter α, obtained via direct 2D WAL fitting of data at $T$=2K or with the background classical magneto-conductivity included.

|  | S1 | S2 | S3 | S4 | S5 |
|---|---|---|---|---|---|
| α (direct WAL fitting) | NA | 6.6±0.2 | 5.1±0.2 | 15.5±0.4 | 0.222±0.003 |
| α (including background from linear MR effect) | NA | 0.8±0.3 | 1.5±0.2 | 2.9±0.2 | 0.075±0.002 |

To correctly account for both WAL and the classical MR effect in the analysis, we fitted our magneto-conductivity data according to $G_{xx}(B)$= $\Delta G_{xx}(B)_{WAL}+G_{xx}(B)_C$, the sum of 2D WAL quantum correction (eq. (1)) and a classical term $G_{xx}(B)_C$ =$1/(R_0+R'\times B)$. The simple expression for the classical magneto-conductivity $G_{xx}(B)_C$ is motivated by the linear MR observed in sample S2, S3, S4 and S5 at $B$>1T (Fig. 2) and neglecting of Hall resistance is justified by the smallness of Hall resistance [31]. Since the classical MR is mixed with the WAL effect at $B$<1T in these samples at low $T$, it is difficult to say if the linear MR persists down to $B$=0 as our expression of $G_{xx}(B)_C$ assumes, so a more complicated expression for $G_{xx}(B)_C$ might be necessary in the low field regime. However, the $G_{xx}(B)_C$ expression we use here serves well to demonstrate the essential point that both classical MR and WAL effects contribute in the analysis of magneto-transport, using a simplest phenomenological model for the classical effect. Moreover, looking at the magneto-resistance of S2 and S5 at somewhat elevated temperatures (e.g. 20K) where the WAL has diminished (Fig. 2(c) and Fig. 2(d)), we see that linear MR is a good approximation for the classical MR term down to low fields. In Fig.4 we plot the classical term of $G_{xx}(B)$ as red dashed lines for S2 and S5. It shows that the magneto-conductance change in S2 is dominated by the classical effect in S2, naturally explaining why the fitting parameter $\alpha$ was too large when the data were interpreted as due to WAL alone. We list in Table 1 the fitting parameter $\alpha$ obtained by fitting the magneto-conductivity data to 2D WAL alone versus taking into account both the WAL and classical effects. After correctly considering the classical MR effect as a background contribution in the fitting of $G_{xx}(B)$ data, the fitting parameter $\alpha$ became significantly smaller and more reasonable, in particular for samples S2-S4 with higher conductivity value and larger classical magneto-conductance change. This insight of classical MR's impact in the low field regime may offer an explanation on why the fitted coefficient $\alpha$ appeared to be too large in some WAL analysis of the linear MR in literature [26]. For metallic sample S1, the MR was totally dominated by the classical effect and thus no meaningful WAL analysis could be performed.

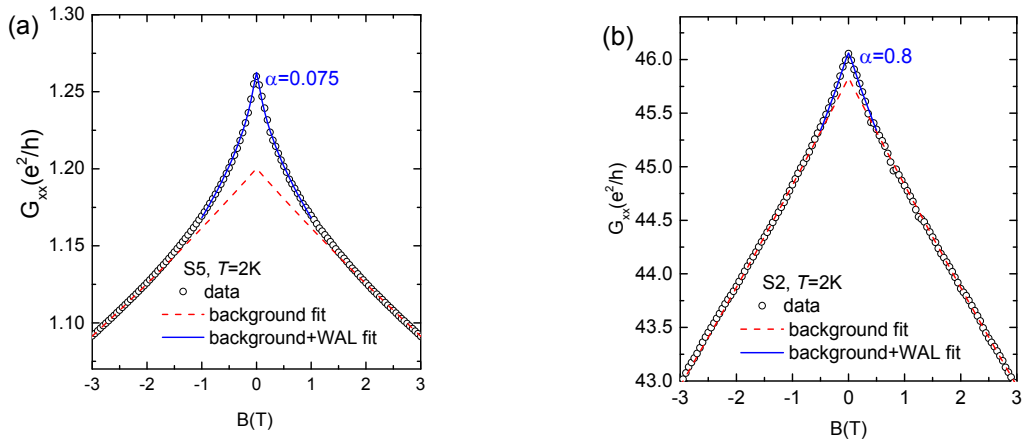

**Figure 4** Fitting (blue solid line) of magneto-conductivity at 2K for sample S5 (a) and S2 (b), using the model of 2D WAL plus the classical background due to linear MR effect. The linear MR induced magneto-conductance background is shown as red dashed line, which accounts for much of the magneto-conductance change, especially in high conductivity sample S2.

## 4 Conclusions

In conclusion, magneto-transport property of $Bi_2Te_3$ films was investigated. While the metallic sample exhibits MR dominated by classical effect [31], samples with less metallic behavior showed a sharp MR dip around $B$=0 at low temperatures due to the weak antilocalization effect. To correctly analyze the magneto-conductivity data using the 2D WAL model, we found that the classical MR induced change in the magneto-conductivity should also be taken into account. Otherwise the 2D WAL fitting alone would yield an overly large coefficient $\alpha$ in the WAL correction. This study provides new insights into the understanding and interpretation of magneto-resistance effects in topological insulators.

## Acknowledgements

X. P. A. G. acknowledges the NSF CAREER Award program (grant # DMR-1151534) for financial support of research at CWRU and the Lee Hsun Young Scientist award of IMR, Chinese Academy of Sciences. Z.D.Z acknowledges the National Natural Science Foundation of China with Grant No. 51331006.